\documentclass[a4paper,11pt]{article}
\usepackage{pos}

\title{Transition magnetic moments for $\Delta \rightarrow p$ transition in  asymmetric nuclear matter}

\author[a]{Suneel Dutt}
\author[a]{Arvind Kumar}
\author*[a]{Harleen Dahiya}

\affiliation[a]{Department of Physics,\\ Dr. B.R. Ambedkar National
	Institute of Technology,\\ Jalandhar, 144008, India}

\emailAdd{dutts@nitj.ac.in}
\emailAdd{kumara@nitj.ac.in}
\emailAdd{dahiyah@nitj.ac.in}

\abstract{In the present work we calculate the transition magnetic moments
for the radiative decays of $\Delta$ baryon to proton $(\Delta \rightarrow p)$ in isospin
asymmetric nuclear medium at finite temperature using chiral SU(3) quark mean field model. Within the framework of chiral
SU(3) mean field model, the properties of baryons in
asymmetric medium are modified through the exchange of scalar fields
$(\sigma, \zeta, \delta)$ and vector fields $(\omega, \rho)$. The isospin asymmetry
of medium is taken into account via scalar-isovector field $\delta$ and vector iso-vector field $\rho$. We calculate the in-medium masses of quarks,
proton and $\Delta$ baryon in asymmetric  matter within the chiral
SU(3) quark mean field model and use these as input in the chiral
constituent quark ($\chi$CQM) model to calculate the in-medium transition magnetic moments for 
$(\Delta \rightarrow p)$ transition for different values of
isospin asymmetry  of hot and dense medium.
For calculating the magnetic moments of baryons, contributions of valence quarks, quark sea and orbital angular momentum of quark sea
are considered in these calculations.
}

\FullConference{The XVIth Quark Confinement and the Hadron Spectrum Conference (QCHSC24)\\
 19-24 August, 2024\\
 Cairns Convention Centre, Cairns, Queensland, Australia\\}

\begin{document}
\maketitle

\section{Introduction}
Exploring the impact of finite density and temperature of hadronic matter on the in-medium properties of various mesons and baryons is of considerable interest. Different experimental programs are aimed to understand different regimes of QCD phase diagram and exploring the
nature of phase transition from hadronic to QGP phase. The experimental facilities such large hadron collider (LHC) \cite{evans2007large, DiNezza:2024ctw}  and relativistic heavy-ion collider (RHIC) \cite{Harrison:2003sb,friman2011cbm,kumar2013star,odyniec2010rhic} where heavy-ion collisions are performed at large center of mass energies 
investigate the QCD phase diagram at high temperature and low baryon densities. Complimentary to this, compressed baryonic matter (CBM) experiment of FAIR project \cite{senger2012compressed,rapp2010charmonium}, NICA project of 
Nuclotron-based Ion Collider Facility (NICA) at Dubna \cite{Kekelidze:2017ghu, Kekelidze:2017tgp} and low energy beam scan programs (BES-I and BES-II) of RHIC  \cite{Odyniec:2019kfh, Liu:2022cqh, Liu:2022wme} aim to investigate 
the QCD phase diagram at high baryon density and
moderate temperature.

Electromagnetic properties of baryons such as their magnetic moments and electromagnetic form factors play significant role in understanding the non-perturbative QCD and exploring the structure of hadrons. EMC effect as observed by  European Muon
Collaboration provided the first clue that the properties of nucleons are modified in the nuclear medium 
\cite{EuropeanMuon:1983wih} and this triggered the interest in understanding the behaviour of hadrons in the medium with finite baryon density.
Various theoretical approaches, for examples, different quark  models \cite{Faessler:2006ft,Schlumpf:1993rm}, QCD sum rules \cite{
Lee:1997jk}, light cone QCD sum rule \cite{Aliev:2008ay,Aliev:2000cy}, chiral perturbation theory \cite{Li:2016ezv},  soliton model \cite{Ledwig:2008es}, chiral constituent quark model ($\chi$CQM) \cite{Dahiya:2002fp,Girdhar:2015gsa} etc. have been used to
explore the magnetic moments and form factors of baryons.
Interest has grown in exploring the electromagnetic properties of different  baryons and mesons in dense hadronic medium  \cite{Ryu:2008st,Ryu:2010tr,Ramalho:2012pu,Arifi:2024tix}.
Apart from the magnetic moments and electromagnetic form factors of given baryons,
the transition magnetic moments also play a key role in revealing the  internal structure and deformation in the structure of the baryons.
The
transition amplitudes for
$\Delta^{+} \rightarrow p \gamma$ transition contain the magnetic dipole $(G_{M1})$, electric quadrupole $(G_{E2})$, and coulomb
quadrupole $(G_{C2})$ contributions and gives the information about the
magnetic dipole moment, electric quadrupole moment and charge quadrupole moment, respectively. The transition magnetic moments of baryons have also been studied in the free space \cite{Li:2017vmq,Fayyazuddin:2020vpe,Kim:2020lgp}.

In the present work, we use the combined approach of chiral SU(3) quark mean field (CQMF) model \cite{Wang:2001jw}
and $\chi$CQM \cite{Dahiya:2018ahb} to calculate the transition magnetic moment of $\Delta^{+} \rightarrow p \gamma$ transition in the isospin asymmetric nuclear medium at finite temperature. 
The magnetic moments of baryons are calculated
using the $\chi$CQM and in-medium effects will be stimulated through CQMF  model \cite{Singh:2016hiw,Singh:2017mxj,Singh:2020nwp,Kumar:2023owb,Dutt:2024lui}.
In the  CQMF  model, quarks are confined inside the baryons through a confining potential and interact through the exchange of scalar meson fields $\sigma,\zeta$ and $\delta$ and the vector fields $\omega$ and $\rho$. The properties of baryons, for example, their in-medium masses are expressed in terms of  properties of 
constituent quark masses. The CQMF model has been used to study the properties of nuclear matter \cite{Wang:2001jw} and neutron stars \cite{Wang:2005vg}.   CQMF model  has also been used along with light quark model to study the pion and kaon structure in asymmetric nuclear matter \cite{Kaur:2024wze,Puhan:2024xdq,Singh:2024lra}.
The $\chi$CQM  considers the contribution of  not only valence quarks, but also sea quarks and orbital angular momentum of sea quarks in calculating the magnetic moments of baryons \cite{Dahiya:2018ahb}.

Following is the outline of present work: In 
Sec. \ref{Sec:formalism} we present the details of
CQMF and $\chi$CQM models. The results on the transition magnetic moments for
$\Delta \rightarrow p$ transition are presented in Sec. \ref{sec:results}.  
\section{Formalism}
\label{Sec:formalism}
As discussed before, in the present manuscript we use the combined approach of CQMF model and $\chi$CQM  to study the in-medium transition magnetic moment.
In Sec. \ref{sec:CQMF} we shall discuss the details of CQMF model to calculate the in-medium masses of quarks and baryons. The 
$\chi$CQM used to calculate the transition magnetic moments of baryons is discussed in  
Sec. \ref{sec:chiCQM}.
\subsection{Chiral SU(3) quark mean field model}
\label{sec:CQMF}
The CQMF model is based on low energy properties of QCD  and incorporates chiral symmetry and its spontaneous and explicit breaking \cite{Wang:2001jw}. As discussed before, the model consider quarks confined inside the baryons as degrees of freedom. The scalar fields $\sigma, \zeta$ and $\delta$
provide the attractive interactions, whereas the vector fields $\omega$ and $\rho$ 
contribute to the repulsive. 
The thermodynamics of asymmetric nuclear medium with finite temperature and density can be studied using the thermodynamic potential
\begin{equation}
\Omega = -\frac{T}
{(2\pi)^3} \sum_{i} \gamma_i
\int_0^\infty d^3k\biggl\{{\rm ln}
\left( 1+e^{- [ E^{\ast}_i(k) - \nu_i^* ]/T}\right) \\
+ {\rm ln}\left( 1+e^{- [ E^{\ast}_i(k)+\nu_i^* ]/T}
\right) \biggr\} -{\cal L}_{M}-{\cal V}_{\text{vac}}.
\label{Eq_therm_pot1}  
\end{equation} 
In above, $i = p,n$, $\gamma_i=2$ is the degeneracy factor and
$E^{\ast }(k)=\sqrt{M_i^{\ast 2}+k^{2}}$. The effective chemical potential $\nu_i^*$ of nucleons relates to free chemical potential $\nu_i$ as \cite{Wang:2001jw}
\begin{align}
\nu_i^* = \nu_i - g_{\omega}^i\omega -g_{\rho}^i I^{3i} \rho.
\end{align}
In Eq. (\ref{Eq_therm_pot1}), the term, ${\cal L}_{M}$ 
consider the contributions from self-interactions of scalar and vector mesons as well as from the explicit symmetry breaking term of the model, i.e., $
{\cal L}_{M} \, = 
{\cal L}_{S} \,+\, {\cal L}_{V} \,+\, {\cal L}_{\chi SB}\,
$. Here, ${\cal L}_{S}$ describes the self-interactions of scalar mesons  and is given as 
\begin{align}
{\cal L}_{S} =& -\frac{1}{2} \, k_0\chi^2
\left(\sigma^2+\zeta^2+\delta^2\right)+k_1 \left(\sigma^2+\zeta^2+\delta^2\right)^2
+k_2\left(\frac{\sigma^4}{2} +\frac{\delta^4}{2}+3\sigma^2\delta^2+\zeta^4\right)\nonumber \\ 
&+k_3\chi\left(\sigma^2-\delta^2\right)\zeta 
 -k_4\chi^4-\frac14\chi^4 {\rm ln}\frac{\chi^4}{\chi_0^4} +
\frac{\xi}
3\chi^4 {\rm ln}\left(\left(\frac{\left(\sigma^2-\delta^2\right)\zeta}{\sigma_0^2\zeta_0}\right)\left(\frac{\chi^3}{\chi_0^3}\right)\right). \label{scalar0}
\end{align}    
In case of vector mesons, within the CQMF, we have the interaction Lagrangian density
\begin{equation}
{\cal L}_{VV}=\frac{1}{2} \, \frac{\chi^2}{\chi_0^2} \left(
m_\omega^2\omega^2+m_\rho^2\rho^2 \right)+g_4\left(\omega^4+6\omega^2\rho^2+\rho^4\right). 
\label{vector}
\end{equation}
The Lagrangian density $
{\cal L}_{\chi SB}$ describing the explicit breaking of chiral symmetry is written as
\begin{equation}
\label{L_SB}
{\cal L}_{\chi SB}=\frac{\chi^2}{\chi_0^2}\left[m_\pi^2f_\pi\sigma +
\left(
\sqrt{2} \, m_K^2f_K-\frac{m_\pi^2}{\sqrt{2}} f_\pi\right)\zeta\right].
\end{equation}

In order to investigate the properties of isospin asymmetric nuclear matter at finite temperature and density we use the mean field approximation \cite{Wang:2001jw}. 
In Eq. (\ref{Eq_therm_pot1}), the term ${\cal V}_{\text{vac}}$
is subtracted to have zero vacuum energy.
The Dirac equation, under the influence of meson mean field, for the quark field $\Psi_{qi}$, is given as 
\begin{equation}
\left[-i\vec{\alpha}\cdot\vec{\nabla}+\chi_c(r)+\beta m_q^*\right]
\Psi_{qi}=e_q^*\Psi_{qi}, \label{Dirac}
\end{equation}
where the subscripts $q$ and $i$ denote the quark $q$ ($q=u, d, s$)
in a baryon of type $i$ ($i=n,p$)\,
and $\vec{\alpha}$\,, $\beta$\, are the usual Dirac matrices.
The effective quark mass $m_{q}^*$ is defined as
\begin{equation}
m_q^*=-g_\sigma^q\sigma - g_\zeta^q\zeta - g_\delta^q I^{3q} \delta + m_{q0}, \label{qmass}
\end{equation}
where $m_{q0}$ is zero for non-strange `$u$' and `$d$' quarks, whereas for strange `$s$' quark $m_{q0}=\Delta m=77$ MeV. Effective energy of particular quark under the influence of meson mean field is given as,
$e_q^*=e_q-g_\omega^q\omega-g_\rho^q I^{3q}\rho \,$ \cite{Wang:2001jw}.
The in-medium mass $M_i^*$ of the baryon is written in term of effective energy $E_i^*$ and spurious center of mass momentum $p_{i \, \text{cm}}^{*}$ \cite{Barik:2013lna}, i.e.,
\begin{align}
M_i^*=\sqrt{E_i^{*2}- <p_{i \, \text{cm}}^{*2}>}. \label{baryonmass}
\end{align} 
In above, $E_i^*$ is related to effective energy of constituent quarks through relation
$ 
E_i^*=\sum_qn_{qi}e_q^*+E_{i \, \text{spin}}$,
where $E_{i \, \text{spin}}$ 
 is fitted to obtain correct masses of baryons in the vacuum.
The thermodynamic potential defined by Eq. (\ref{Eq_therm_pot1}) is minimized with respect to the scalar fields
 $\sigma$, $\zeta$ and $\delta$,
 the dilaton field, $\chi$,  and, the vector fields $\omega$ and $\rho$, i.e.,
 \begin{align}
  \frac{\partial \Omega}{\partial \sigma} = 
  \frac{\partial \Omega}{\partial \zeta} =
  \frac{\partial \Omega}{\partial \delta} =
  \frac{\partial \Omega}{\partial \chi} =
  \frac{\partial \Omega}{\partial \omega} =
  \frac{\partial \Omega}{\partial \rho} = 
    0.
    \label{eq:therm_min1}
  \end{align}
  The coupled system of non-linear equations
  as 
  obtained above are solved for different values of baryon density, temperature and isospin asymmetry of the medium. We define the
 isospin asymmetry in the medium through $I_a = -\frac{\Sigma_i I_{3i} \rho_{i}}{\rho_{B}}$.
  Here  $I_{3i}$ denotes the 3$^{rd}$ component of isospin quantum number in the $i^{th}$ baryon and $\rho_B$ is the total baryonic density.

\subsection{Chiral constituent quark model: $\Delta \rightarrow p$ transition magnetic moment}
\label{sec:chiCQM}
To obtain the magnetic moment for $\Delta \rightarrow p$ transition contributions of 
valence quarks, sea quarks and orbital angular momentum of sea quarks are taken into account.
In $\chi$CQM the internal constituent quark undergoes the emission of Goldstone boson and this
further splits into the pair of quark-antiquark.
\cite{Cheng:1994zn}.
The Lagrangian density  ${\cal L} = g_8 \bar{q} \Phi q$ describes the interactions of quarks with Goldstone bosons, where  $\Phi$ consists of  octet of pseudoscalar mesons as well as pseudoscalar singlet $\eta^{'}$, i.e., 
\begin{equation}
\Phi=\left(\begin{array}{ccc}\frac{\pi^o}{\sqrt{2}}+\beta \frac{\eta}{\sqrt{6}}+\zeta^{'} \frac{\eta^{\prime}}{4\sqrt{3}} & \pi^{+} & \alpha K^{+} \\ \pi^{-} & -\frac{\pi^o}{\sqrt{2}}+\beta \frac{\eta}{\sqrt{6}}+\zeta^{'} \frac{\eta^{\prime}}{4\sqrt{3}} & \alpha K^0 \\ \alpha K^{-} & \alpha \bar{K}^0 & -\beta \frac{2 \eta}{\sqrt{6}}+\zeta^{'} \frac{\eta^{\prime}}{4\sqrt{3}}\end{array}\right).
\end{equation}
The constraints  $m_s>m_{u, d}$ and also  non-degenerate masses of pseudoscalars, i.e., $\left(M_{\eta^{\prime}}>M_{K, \eta}>M_\pi\right)$ incorporate SU(3) symmetry breaking. 
Also, parameter $a=\left|g_{8}\right|^2$ denotes  the transition probability of chiral fluctuation of the splitting  $u(d) \rightarrow$ $d(u)+\pi^{+(-)}$. Further,  $a \alpha^2, a \beta^2$ and $a \zeta^{'2}$ represent the probabilities of transitions of $u(d) \rightarrow s+K^{-(0)}, u(d, s) \rightarrow$ $u(d, s)+\eta$ and $ u(d, s) \rightarrow u(d, s)+\eta^{\prime}$, respectively
 \cite{Girdhar:2015gsa,Sharma:2010vv}.

Total transition magnetic moment for 
$\Delta \rightarrow p$ transition in asymmetric nuclear medium is written as
\cite{Dahiya:2018ahb}

\begin{align}
\mu^{*} \left(\Delta_{\frac{3}{2}^{+} \rightarrow \frac{1}{2}^{+}}\right) =
\mu^{*} \left(\Delta_{\frac{3}{2}^{+} \rightarrow \frac{1}{2}^{+}}\right)_V +
\mu^{*} \left(\Delta_{\frac{3}{2}^{+} \rightarrow \frac{1}{2}^{+}}\right)_S + 
\mu^{*} \left(\Delta_{\frac{3}{2}^{+} \rightarrow \frac{1}{2}^{+}}\right)_O.
\end{align}
The individual contributions to the total magnetic moment are expressed in terms of quark magnetic moments $\mu_q^*$ and spin polarization $\Delta q$ as
\begin{align}
\mu^{*} \left(\Delta_{\frac{3}{2}^{+} \rightarrow \frac{1}{2}^{+}}\right)_V &= 
\sum_{q = u,d,s} \Delta q \left(\Delta_{\frac{3}{2}^{+} \rightarrow \frac{1}{2}^{+}}\right)_V \mu_q^{*} \nonumber\\
\mu^{*} \left(\Delta_{\frac{3}{2}^{+} \rightarrow \frac{1}{2}^{+}}\right)_S &= 
\sum_{q = u,d,s} \Delta q \left(\Delta_{\frac{3}{2}^{+} \rightarrow \frac{1}{2}^{+}}\right)_S \mu_q^{*}\nonumber\\
\mu^{*} \left(\Delta_{\frac{3}{2}^{+} \rightarrow \frac{1}{2}^{+}}\right)_O &= 
\sum_{q = u,d,s} \Delta q \left(\Delta_{\frac{3}{2}^{+} \rightarrow \frac{1}{2}^{+}}\right)_V 
\mu^{*} \left(q_{+} \rightarrow \right).
\end{align}
The nuclear medium effects are introduced in the above equation through the  effective quark magnetic $\mu_q^{*}$, which individually for $u,d $ and $s$ quarks are
defined as
	\begin{equation}
 	 \mu_d^* =-\left(1-\frac{\Delta M}{M_B^*}\right),~~  \mu_s^*=-\frac{m_u^*}{m_s^*}\left(1-\frac{\Delta M}{M_B^*}\right),~~  \mu_u^*=-2\mu_d^* .\label{magandmass}         
 	\end{equation}
 	The above definitions for the
calculations of magnetic moments of quarks are motivated to incorporate the confinement effects \cite{Sharma:2010vv}. Here, $\Delta M = M_{vac} - M_B^{*}$ and also, $M_{vac}$ and $M_B^{*}$ are the vacuum and  in-medium masses of initial state baryon. The effective mass of baryons is calculated using Eq.(\ref{baryonmass}).
The factor $\mu^{*} \left(q_{+} \rightarrow \right)$ consider the contribution of orbital moment of chiral fluctuations and is expressed in terms of transition probabilities and masses of Goldstone bosons.
In Ref. \cite{Dahiya:2018ahb}, the transition magnetic moment for
$\Delta \rightarrow p$ transition are calculated in the free space.

 
 \begin{figure}
    \centering
    \includegraphics[width=0.9\linewidth]{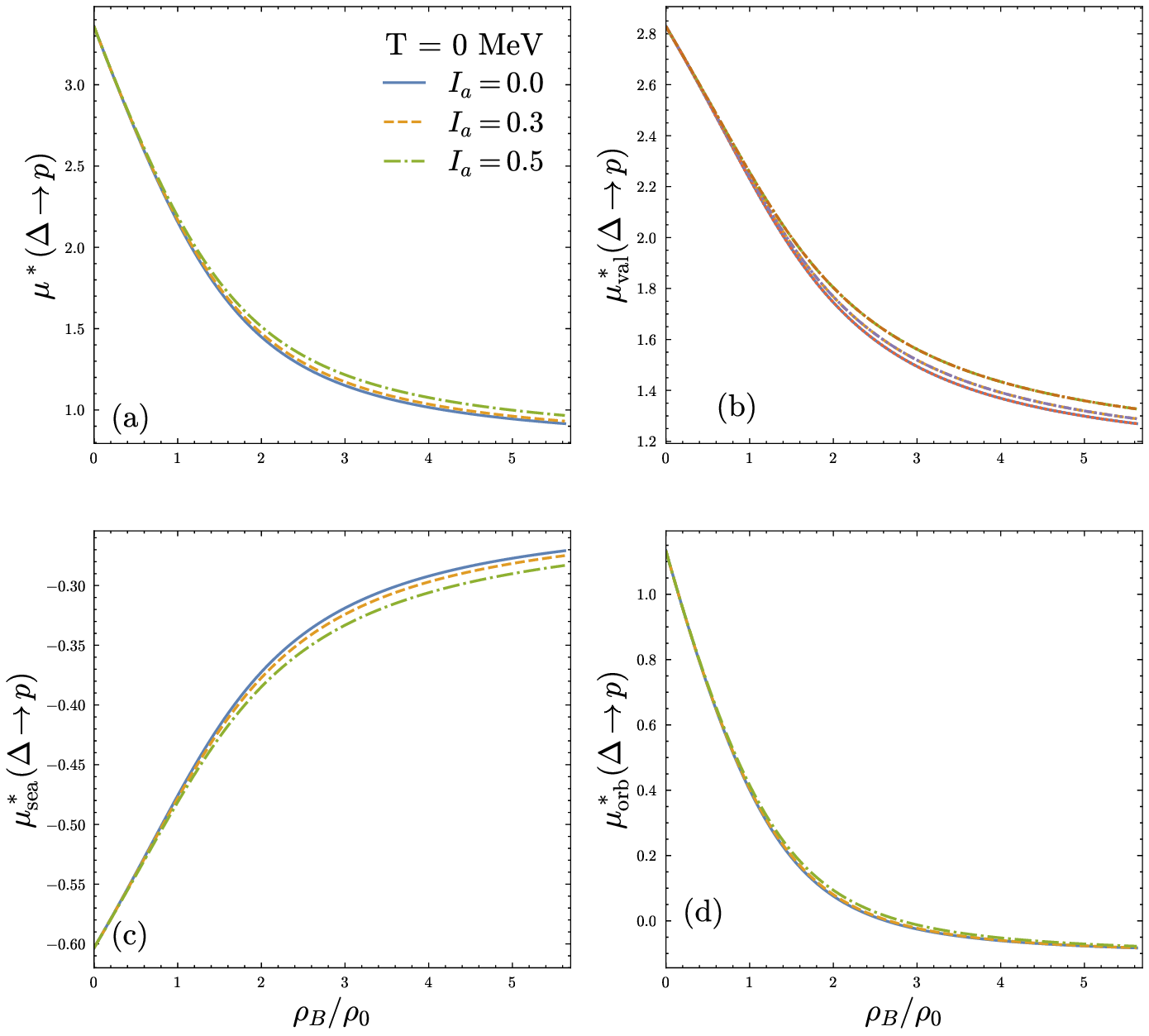}
    \caption{In above, (a) transition magnetic moment for $\Delta \rightarrow p$, (b) contribution of valence quarks, (c) sea quarks and (d) orbital angular momentum of sea quarks is shown as a function of baryon density $\rho_B$ (in units of $\rho_0$) of nuclear matter at temperature $T = 0$ MeV and isospin asymmetry $I_a = 0, 0.3$ and 0.5.}
    \label{fig_Delta}
\end{figure}

\section{Results and discussion}
\label{sec:results}
In this section, we discuss the results on the in-medium transition magnetic moments for $\Delta \rightarrow p$ transition calculated using the $\chi$CQM and incorporating the medium effects through the CQMF model. As described previously, the effective constituent quark masses and masses of baryons are calculated using the CQMF model and are used as input in $\chi$CQM.
The isospin asymmetric effects are introduced through the asymmetry parameter $I_a$. For a given asymmetry and temperature of the medium, the effective quark masses are observed to decrease as the baryon density is increased. The decrease in the mass of quarks becomes slightly slower as value of $I_a$ is increased from zero to finite value.
Finite value of $I_a$ also causes  the mass splitting between the masses of $u$ and $d$ quarks due to non-zero value of scalar-isovector $\delta$ field in asymmetric nuclear medium. The details about the magnitude of changes in effective masses of constituent quarks and baryons in asymmetric nuclear medium  can be found in Refs. \cite{Singh:2016hiw,Singh:2017mxj,Singh:2020nwp}. In the current work, we present the results on transition magnetic moments.
In Fig. \ref{fig_Delta}a, we plot the net transition magnetic moment for 
$\Delta \rightarrow p$ transition as a function of baryons density ratio $\rho_B/\rho_0$. The results are shown for temperature $T =0$ and isospin asymmetry parameter $I_a = 0, 0.3$ and $0.5$. 
As can be seen, an increase in the baryon density of medium causes a decrease in the value of $\mu^{*}\left(\Delta \rightarrow p\right)$. 
For a given density, increasing the value of $I_a$ from zero to $0.5$ causes an increase in $\mu^{*}\left(\Delta \rightarrow p\right)$. The impact of isospin asymmetry is found to be more appreciable at large baryon density of the medium.
In symmetric nuclear medium ($I_a = 0$), as baryon density is increased from zero to $\rho_0$ (3$\rho_0$), the value of
$\mu^{*}\left(\Delta \rightarrow p\right)$
decreases by $35.71\%(65.77\%)$.
Increasing $I_a$ to 0.5, these values at
$\rho_0$ (3$\rho_0$) change to
 $34.82\%(63.69\%)$. 
For better understanding, in
Fig. \ref{fig_Delta}(b), (c) and (d),
the individual contributions to total magnetic moment from
valence quarks, sea quarks and orbital angular momentum of sea quarks, respectively, are shown. The magnitude of all three contribution decrease as a function of density of the medium. As a function of $I_a$, magnetic moment contributions from valence quarks increase, whereas from sea quarks become more negative. In terms of magnitude, changes are observed to be largest in case of valence quark, i.e.,
Fig. \ref{fig_Delta}b.
At temperature $T = 100$ MeV, the values of magnetic moment, as $\rho_B$ changes from vacuum to $\rho_0 (3\rho_0)$, decreases by $30.65\% (61.31 \%)$ at $I_a = 0$, whereas
at $I_a = 0.5$ these values change to
$30.36\%$ (60.42\%).
The present work of in-medium transition magnetic moments will be extended to other transitions from decuplet to octet baryons 
 in future work.

 \section*{Acknowledgment}
 A. K. sincerely acknowledge 
Anusandhan National Research Foundation (ANRF),
Government of India for funding of the research project under the
Science and Engineering
Research Board-Core Research Grant (SERB-CRG) scheme (File No. CRG/2023/000557).
H.D. would like to thank  the Science and Engineering Research Board, Anusandhan-National Research Foundation (ANRF), Government of India under the SERB-POWER Fellowship scheme (Ref No. SPF/2023/000116) for financial support.


\bibliographystyle{elsarticle-num} 

\bibliography{Ref}

%

\end{document}